\documentclass[pra,showpacs,showkeys,floatfix,twocolumn,superscriptaddress]{revtex4}

\usepackage{graphicx}\usepackage{units}\usepackage{epsfig}\usepackage{xspace}\usepackage[american]{babel}\usepackage{color}

\begin{document} 

\title{Expansion dynamics of a dipolar Bose-Einstein condensate}

\author{S.\ Giovanazzi} 	
\affiliation{5.\ Physikalisches Institut, Universit\"at Stuttgart, Pfaffenwaldring 57, 70550 Stuttgart, Germany} 
\author{P.\ Pedri} 		\affiliation{LPTMS, B\^{a}timent 100, Universit\'{e} Paris-Sud, 91405 ORSAY cedex, France}
\author{L.\ Santos} 		\affiliation{Institut f\"ur Theoretische Physik III, Universit\"at Stuttgart, Pfaffenwaldring 57, 70550 Stuttgart, Germany}
\author{A.\ Griesmaier} 	\affiliation{5.\ Physikalisches Institut, Universit\"at Stuttgart, Pfaffenwaldring 57, 70550 Stuttgart, Germany}
\author{M.\ Fattori}		\affiliation{5.\ Physikalisches Institut, Universit\"at Stuttgart, Pfaffenwaldring 57, 70550 Stuttgart, Germany}
\author{T.\ Koch}		\affiliation{5.\ Physikalisches Institut, Universit\"at Stuttgart, Pfaffenwaldring 57, 70550 Stuttgart, Germany}
\author{J.\ Stuhler} 		\affiliation{5.\ Physikalisches Institut, Universit\"at Stuttgart, Pfaffenwaldring 57, 70550 Stuttgart, Germany}
\author{T.\ Pfau}		\affiliation{5.\ Physikalisches Institut, Universit\"at Stuttgart, Pfaffenwaldring 57, 70550 Stuttgart, Germany}

\date{\today}

\begin{abstract} Our recent measurements on the expansion of a chromium dipolar condensate after release from an optical trapping potential are in good agreement with an exact solution of the hydrodynamic equations for dipolar Bose gases. We report here the theoretical method used to interpret the measurement data as well as more details of the experiment and its analysis. The theory reported here is a tool for the investigation of different dynamical situations in time-dependent harmonic traps. \end{abstract} 
 \maketitle

\section{Introduction}

Recent developments in the manipulation of cold atoms and molecules are paving the way towards the analysis of polar gases, for which the dipole-dipole interaction may play a significant, or even dominant role.   In this sense, exciting perspectives towards the generation of ultra cold polar molecules have been recently opened by experiments on direct cooling and trapping of molecules, as well as on photoassociation and on Feshbach resonances in binary mixtures of ultra cold atoms \cite{doyle:2004,Sage:2005,Stan:2004,Inouye:2004,Simoni:2003b,Meerakker:2005}. However, up to now, no degenerate gas of ultra cold polar molecules has been ever realized.

On the other hand, the realization of a Bose-Einstein condensate (BEC) of polarized chromium \cite{Griesmaier:2005a} constitutes the first example of a quantum degenerated dipolar gas. The large magnetic moment of chromium ($6$ Bohr magnetons)  makes its magnetic dipole-dipole interaction sufficiently strong to induce qualitative differences in the BEC properties.

Dipole-dipole interactions are long-range and anisotropic (partially attractive and partially repulsive), in clear contrast to the up to now usual short-range isotropic interactions. Chromium is hence the first atomic species to be Bose-condensed that has a visible anisotropic interaction. By means of an appropriate rotating magnetic field, this anisotropy may be employed to tune the dipolar interactions \cite{Giovanazzi:2002a}, introducing a second control mechanism in addition to the tuning of the isotropic interactions by means of Feshbach resonances \cite{Tiesinga:1993}. The dipolar anisotropy should also cause sound to propagate with different speed in different directions providing an interesting tool to investigate dissipation mechanisms, e.g. the Landau criterion for superfluidity.

Moreover, the partially attractive and partially repulsive nature of the dipolar interaction together with its long range character makes the question of stability for strong dipolar interactions more intricate \cite{Goral:2000b, Santos:2000a, Giovanazzi:2004a}. Indeed, an homogeneous dipolar condensate is unstable when the dipolar interaction is stronger than the s-wave scattering interaction \cite{Goral:2000b,Santos:2000a,Martikainen:2001a}, an issue which may become especially relevant for ultra cold heteronuclear molecules with electric dipole moments of the order of one Debye.

Many other new exciting phenomena are expected in dipolar quantum gases oriented by an external field. Recent theoretical analyses have shown that stability and excitations of dipolar gases are crucially determined by the trap geometry \cite{Goral:2000b,Yi:2000a,Santos:2000a,Yi:2002,Goral:2002,Baranov:2002b,ODell:2004a,Santos_Roton03}. Dipolar degenerate quantum gases are also attractive in the context of strongly-correlated atoms \cite{Goral:2002a,Baranov:2005,Rezayi:2005,Barnett:2006,Micheli:2006}, as physical implementation of quantum information \cite{DeMille:2002a,Jaksch:2000} and for the study of ultracold chemistry \cite{Bodo:2002}.

In a recent Letter \cite{Stuhler:2005}, we reported the first observation of mechanical effects due to the magnetic dipole-dipole interaction in a degenerate quantum gas. We investigated the expansion of a chromium BEC polarized by an external magnetic field after release from an anisotropic trap. The anisotropy of the magnetic dipolar interaction was shown to lead to a measurable anisotropic deformation of the expanding chromium BEC, which is quantitatively in very good agreement with the theoretical predictions. The expansion technique has been used since the earliest experiments with cold atoms in order to show the existence of the Bose-Einstein condensed phase \cite{Anderson:1995a,Castin1996a} and to probe the coherence properties of bosonic atoms on a lattice \cite{Greiner:2002a} and in the context of ultra cold Fermi gases to point out superfluid effects \cite{OHara:2002a,Menotti:2002}.
In this paper, we report in detail the theoretical methods that have been used to interpret the experimental data on the expansion, as well as a more detailed description of the experiment. The theory generalizes a recent exact result obtained for the Thomas-Fermi dynamics of a dipolar condensate \cite{ODell:2004a, Eberlein:2005} obtained explicitly for the case of cylindrical-symmetric traps  to the case of non-axisymmetric traps.

The paper is organized as follows: In section 2.A we introduce the hydrodynamic theory of a dipolar superfluid. The general equations for the dynamics of dipolar condensate in the hydrodynamic limit in the presence of a general time-dependent non-axisymmetric  harmonic trap are presented in section 2.B. The calculation of the ground state density profile and of the expansion of a dipolar condensate are a direct application of the theory (sections 2.C and 2.D, respectively), which are used to explain the mechanisms that are responsible for the reported observations quantitatively and qualitatively. In section 3.A we briefly summarize the experimental procedure we used to obtain the expansion data, which are compared with the theory subsequently in section 3.B. Section 4 conclude. Appendix A discusses some properties of the dipole-dipole mean-field integral. Appendix B contains the expression for the mean-field dipole-dipole potential.

\section{Theory}

The expansion of the chromium Bose-Einstein condensate can be studied theoretically by solving the generalized form of the time-dependent Gross-Pitaevskii equations given in Subsection 2.A. These equations are a tool to investigate a large number of dynamical situations in addition to the expansion, e.g. large amplitude collective oscillations (see for instance the review \cite{BEC_review} for the case of pure s-wave contact interaction)  and their frequencies \cite{Giovanazzi:2006c}. The generalized time-dependent Gross-Pitaevskii equations (\ref{ggpe}) and (\ref{mf}) or the equivalent hydrodynamic equations (\ref{continuity}) and (\ref{euler}) have been introduced earlier to study the static and stability properties of a dipolar condensate theoretically in \cite{Goral:2000b, Santos:2000a, LiYou:2003} for the case of magnetic dipole-dipole interaction and in \cite{Odell:2000a, Giovanazzi:2001a, Giovanazzi:2001b} for the case of laser-induced dipole-dipole interaction. The expansion of a dipolar condensate has been previously theoretically investigated in the case of cylindrical symmetry in \cite{Giovanazzi:2003a} for the Thomas-Fermi limit and in \cite{LiYou:2003,Goral:2002} for the case of an s-wave scattering length tuned close to zero.

\subsection{Hydrodynamic equations of a dipolar superfluid}

The generalized time-dependent Gross-Pitaevskii equation is given by
\begin{eqnarray}\label{ggpe} - i \hbar \frac{\partial}{\partial t} \Psi(\vec{r},t) &=& -\frac{\hbar^2}{2m}\nabla^2 \,\Psi(\vec{r},t) + V_{\rm ext}(\vec{r},t)\Psi(\vec{r},t)   \nonumber\\ &\,&+ V_{\rm mf}(\vec{r},t)\,\Psi(\vec{r},t) \end{eqnarray}
where $V_{\rm ext}$ is an external potential and $V_{\rm mf}$ is the mean field potential given by
\begin{equation} \label{mf} V_{\rm mf}(\vec{r},t) = g \, n(\vec{r},t) +  \int d^3r' U_{\rm dd}(\vec{r}-\vec{r}\,') n(\vec{r}\,',t) \end{equation}
where $n(\vec{r},t)=|\Psi(\vec{r},t)|^2 $ is the condensate atomic density and $m$ is the atomic mass. In equation (\ref{mf}) $g$ is the s-wave scattering coupling constant given by \begin{equation}\label{g} g = \frac{4 \pi \hbar^2 \,a }{m} \end{equation} where $a$ is the s-wave scattering length. Here \begin{equation}\label{udd} U_{\rm dd}(\vec{r})=\frac{\textstyle \mu_0 \mu^2_{\rm m}}{\textstyle 4 \pi r^3} \left( 1- \frac{\textstyle 3 (\hat{e}_\mu\vec{r})^2}{\textstyle r^2}\right) \end{equation}
is the (dipole-dipole) interaction energy between two equally oriented magnetic dipoles
$\vec{\mu}_{\rm m}=\mu_{\rm m} \hat{e}_\mu$
aligned by a polarizing magnetic field ($\hat{e}_\mu \| \vec{B}$) and with relative coordinate $\vec{r}$.
The polarizing magnetic field $ \vec{B}$ is assumed parallel to one of the symmetry axes of the harmonic trap. A measure of the strength of the dipole-dipole interaction relative to the s-wave scattering energy is given by the dimensionless quantity
\begin{eqnarray} \varepsilon_{\mathrm{dd}} = \frac{\mu_0 \mu^2 m}{12 \pi \hbar^2 a} \label{epsilon} \end{eqnarray}

In complete equivalence of equations (\ref{ggpe}) and (\ref{mf}), one can solve the corresponding collisionless hydrodynamic equations, i.e. the continuity and Euler equations, given respectively by
\begin{eqnarray} {\partial n \over \partial t} &=& - {\vec{ \nabla}} \cdot (n \vec{v}) \label{continuity}\\
m{\partial {\vec{ v}} \over \partial t} &=&-  {\vec{ \nabla}} \left(-\frac{\hbar^2\,\nabla^2 \sqrt{n}}{2 \,m\,\sqrt{n}}  +\frac { m v^2}{2} + V_{\mathrm{ext}} + V_{\mathrm{mf}}\right) \label{euler} \end{eqnarray}
where $\vec{v}$ is the superfluid velocity, which is related to the phase of the macroscopic condensate wave function $\Psi(\vec{r},t)=\sqrt{n(\vec{r},t)} \exp [\mathrm{i} \phi(\vec{r},t)]$ by $\vec{v}(\vec{r},t)=(\hbar/m) \vec{\nabla} \phi(\vec{r},t)$. Eqs.\  (\ref{continuity}) and (\ref{euler}) describe the potential flow of a fluid in the presence of a self-consistent potential due to the presence of the long-range dipolar interaction (the second contribution on the right side of (\ref{mf})) and whose pressure $P$ and density $n$ are related by the equation of state  $P=(g/2) n^2$.

\subsection{Hydrodynamic solutions for time-dependent harmonic potentials}

An exact class of solutions of the generalized GPE  equations (\ref{ggpe}) and (\ref{mf}) or equivalently of the hydrodynamic equations (\ref{continuity}) and (\ref{euler}) in the Thomas-Fermi limit has been obtained for harmonic time-dependent potential \cite{ODell:2004a} of the form given by
\begin{eqnarray} V_{\rm ho}(\vec{r},t)=\frac{m}{2}\left(\omega_x^2(t) x^2+\omega_y^2(t)  y^2+\omega_z^2(t) z^2\right) \end{eqnarray}

In the Thomas-Fermi or hydrodynamic limit the quantum pressure term $\hbar^2\,\nabla^2 \sqrt{n} / (2 \,m\,\sqrt{n}) $ proportional to the Laplacian of the modulus of the wave function is neglected. The solutions have the form given by 
\begin{eqnarray}
n(\vec{r},t) &=& \frac{15 N }{8\pi  R_x R_y R_z} \left[1-\frac{x^2}{R_x^2}-\frac{y^2}{R_y^2} - \frac{z^2}{R_z^2}\right] \label{parabola} \\
\vec{v}(\vec{r},t) & = & \frac{1}{2} \vec{\nabla} \left[ \alpha_x x^{2} + \alpha_y y^{2}+ \alpha_z z^{2} \right]
\label{velocity} 
\end{eqnarray}
valid until the right hand side of (\ref{parabola}) is $ \geq0$ otherwise $n(\vec{r},t)=0$. The existence of this class of solutions for harmonic traps is due to the harmonic nature of the external and self-consistent potentials and the Bernoulli term. The density $n(\vec{r},t)$ and the velocity field $\vec{v}(\vec{r},t)$ depend on time only through the time-dependence of the condensate radii $R_{j}(t)$ and the $\alpha_{j}(t)$ coefficients. The later are simply given by \begin{equation} \alpha_{j}(t)= \frac{\partial}{\partial t}\log(R_{j}(t)) \end{equation} 
The time dependence of the condensate radii $R_{j}$ are given by solving a simpler equation which can be written in a compact form as
\begin{equation}\label{ne}
{N m \over 7} {\mathrm{d}^{2}\,R_{j}\over\mathrm{d}t\,^{2}} = -   {\partial\over\partial
R_{j}} H_{\mathrm{tot}}\left(R_x,R_y,R_z\right)
\end{equation}
where $H_{\mathrm{tot}}/N$ is the expectation value of the total energy per particle. The various contributions to $H_{\mathrm{tot}}$ are: 
\newline (a) the classical contribution to the kinetic energy 
\begin{equation} H_{\mathrm{kin}}= {N m\over14} \left(\dot{R}_x^2 + \dot{R}_y^2 + \dot{R}_z^2\right) \label{hkin} \end{equation}
which does not depend on the radii but only on their time derivative\newline
(b) the potential energy in the harmonic trap
\begin{equation}\label{ho} H_{\mathrm{ho}}={N m\over14} \left(\omega_x^2R_x^2+\omega_y^2R_y^2 + \omega_z^2R_z^2\right) \end{equation}
 (c) the mean-field energy due to s-wave scattering
\begin{eqnarray}\label{hs} H_{s} &=&{15\over 7} \left({ N^2 \hbar^2 a \over  m }\right) {1\over R_x R_{y} R_z} \end{eqnarray}
(d) the mean-field magnetic dipole-dipole energy 
\begin{eqnarray} H_{dd}^{x}&=& - {15\over 7} \left({ N^2 \hbar^2 a \over  m }\right) {\varepsilon_{dd}\,f(\kappa_{yx},\kappa_{zx})\over R_x R_{y} R_z} \label{hddx} \end{eqnarray}
Here the magnetic field is assumed parallel to the $\hat{x}$ axes and $\kappa_{yx}=R_y/R_x$ and $\kappa_{zx}=R_z/R_x$  are condensate aspect ratios. The function $f$ is given by
\begin{eqnarray}\label{ffunction} f(\kappa_{yx},\kappa_{zx})&=&1+3\kappa_{yx}\kappa_{zx}  {\mathrm{E}(\varphi\setminus\alpha) -\mathrm{F}(\varphi\setminus\alpha) \over (1-\kappa_{zx}^{2})\sqrt{1-\kappa_{yx}^{2}}} \end{eqnarray}
with \begin{eqnarray} \sin\varphi&=&\sqrt{1-\kappa_{yx}^{2}}\\ \sin^{2}\alpha&=&{1-\kappa_{zx}^{2}\over1-\kappa_{yx}^{2}} \end{eqnarray}
Here $\mathrm{F}(\varphi\setminus\alpha)$  and $\mathrm{E}(\varphi\setminus\alpha)$ are the incomplete elliptic integrals of the first and second kinds \cite{Abramowitz:}. The function $f$ is a smooth and decreasing function with values in the interval $(-2,1)$. See Figures and discussion on $f$ in appendix A.

\subsection{Equilibrium configuration and scaling property}

The equations for the equilibrium values of the condensate radii can be easily derived from Newton's equations (\ref{ne}) and from equations (\ref{ho}), (\ref{hs}) and (\ref{hddx}) and are given by
\begin{widetext}
\begin{eqnarray}  \label{statics} 
 \omega_x^{2}&=& \left({ 15 \, N \hbar^2 a \over  \, m^{2} \, (R_x)^{3} R_{y} R_z }\right)
\left[ 1-  \varepsilon_{dd}\, f\left({ R_{y} \over R_x}, { R_{z} \over R_x}\right)
+\varepsilon_{dd}\, { R_{y} \over R_x} {\partial f \over \partial \kappa_{1}}\left({ R_{y} \over R_x}, { R_{z} \over R_x}\right)
+\varepsilon_{dd}\, { R_{z} \over R_x} {\partial f \over \partial \kappa_{2}}\left({ R_{y} \over R_x}, { R_{z} \over R_x}\right)
\right] \label{equi1}\\
 \omega_y^{2}&=& \left({ 15 \, N \hbar^2 a \over   \, m^{2} \, R_x (R_y)^{3} R_z }\right)
\left[ 1-  \varepsilon_{dd}\, f\left({ R_{y} \over R_x}, { R_{z} \over R_x}\right)
-\varepsilon_{dd}\, { R_{y} \over R_x} {\partial f \over \partial \kappa_{1}}\left({ R_{y} \over R_x}, { R_{z} \over R_x}\right)\right]
\label{equi2}\\
 \omega_z^{2}&=& \left({ 15 \, N \hbar^2 a \over   \, m^{2} \, R_x R_{y} (R_z)^{3} }\right)
\left[ 1-  \varepsilon_{dd}\, f\left({ R_{y} \over R_x}, { R_{z} \over R_x}\right)
-\varepsilon_{dd}\, { R_{z} \over R_x} {\partial f \over \partial \kappa_{2}}\left({ R_{y} \over R_x}, { R_{z} \over R_x}\right)\right]
\label{equi3}\end{eqnarray}
\end{widetext}

These equations are solved numerically for the case of a trapped Cr condensate. 
Using the value of the chromium scattering length obtained from the Feshbach resonance measurements \cite{Werner:2005}, $\epsilon_{dd}$ results  of order of 
$\epsilon_{dd}^{Cr}\approx 0.15 $
To get insight into the modifications of the condensate radii and of the shape and volume of the condensate, we consider the case of small $\epsilon_{dd}$,  which is also relevant for the chromium condensate. 
The variations of the condensate radii $\Delta R_{i} = R_{i} - R_{i}^{0}$ with $i=x,y,z$ and $R_{i}^{0}= \left(15 \,N \, \hbar^{2} \,a / m^{2}\,\omega_{i}^{2} \right)^{1/5}$  are given to the first order in $\epsilon_{dd}$ by
\begin{widetext}
\begin{eqnarray}  \label{scaledstatics} 
{\Delta R_{x} \over R_{x}^{0}}
&=& -  \varepsilon_{dd} \frac15 f\left({ \omega_{x} \over \omega_y}, { \omega_{x} \over \omega_z}\right)
-  \varepsilon_{dd}\frac12 { \omega_{x} \over \omega_y} {\partial f \over \partial \kappa_{1}}\left({ \omega_{x} \over \omega_y}, { \omega_{x} \over \omega_z}\right)
-  \varepsilon_{dd}\frac12 { \omega_{x} \over \omega_z} {\partial f \over \partial \kappa_{2}}\left({ \omega_{x} \over \omega_y}, { \omega_{x} \over \omega_z}\right)
\label{delta1}\\
{\Delta R_{y} \over R_{y}^{0}} &=&
-  \varepsilon_{dd} \frac15 f\left({ \omega_{x} \over \omega_y}, { \omega_{x} \over \omega_z}\right)
+ \varepsilon_{dd}\frac12 { \omega_{x} \over \omega_y} {\partial f \over \partial \kappa_{1}}\left({ \omega_{x} \over \omega_y}, { \omega_{x} \over \omega_z}\right) \label{delta2}\\ {\Delta R_{z} \over R_{z}^{0}}
&=& -  \varepsilon_{dd} \frac15 f\left({ \omega_{x} \over \omega_y}, { \omega_{x} \over \omega_z}\right) +  \varepsilon_{dd}\frac12 { \omega_{x} \over \omega_z} {\partial f \over \partial \kappa_{2}}\left({ \omega_{x} \over \omega_y}, { \omega_{x} \over \omega_z}\right)
\label{delta3} \end{eqnarray}
\end{widetext}
These can be used to show that the rate of change of the condensate volume, i.e. the product of $V=(4\pi/3)\, R_{x}\,R_{y}\,R_{z}$ is proportional to the function $f$, i.e.
\begin{widetext}
\begin{eqnarray} {\Delta V \over V^{0}} = -  \varepsilon_{dd} \frac3 5 f\left({ \omega_{x} \over \omega_y}, { \omega_{x} \over \omega_z}\right)\label{dv}\end{eqnarray}
In contrast, the rate of change of the aspect ratio is related to the derivative of $f$
\begin{eqnarray}
{\Delta (R_{y}/R_x) \over (R_{y}^{0}/R_x^{0})}&=& \varepsilon_{dd} { \omega_{x} \over \omega_y} {\partial f \over \partial \kappa_{1}}\left({ \omega_{x} \over \omega_y}, { \omega_{x} \over \omega_z}\right)
+  \varepsilon_{dd}\frac12 { \omega_{x} \over \omega_z} {\partial f \over \partial \kappa_{2}}\left({ \omega_{x} \over \omega_y}, { \omega_{x} \over \omega_z}\right)
\label{dk1}\\
{\Delta (R_{z}/R_x) \over (R_{z}^{0}/R_x^{0})}&=& \varepsilon_{dd} \frac12{ \omega_{x} \over \omega_y} {\partial f \over \partial \kappa_{1}}\left({ \omega_{x} \over \omega_y}, { \omega_{x} \over \omega_z}\right)
+  \varepsilon_{dd} { \omega_{x} \over \omega_z} {\partial f \over \partial \kappa_{2}}\left({ \omega_{x} \over \omega_y}, { \omega_{x} \over \omega_z}\right)\label{dk2}
\end{eqnarray}
\end{widetext}
Since the derivatives of $f$ are always negative, the aspect ratios $R_{y}/R_x$ and $R_{z}/R_x$ are always decreasing meaning that {\it the condensate tends to have a shape elongated in the direction of the magnetic field}.
It has been shown numerically that this result is valid also for any value of $\epsilon_{dd}$ in the interval $(0,1)$ (where the Thomas-Fermi approximation has still a clear meaning) and in the case of cylindrical symmetry \cite{Eberlein:2005}. 
Equations (\ref{dv}), (\ref{dk1}) and (\ref{dk2}) are the generalization of equations (16) and (15) of \cite{Giovanazzi:2003a}, which hold in the case of cylindrical symmetry. Equation (16) of \cite{Giovanazzi:2003a} contains an extra term proportional to the derivative of $f$, which is rectified by (\ref{dv}).

It is noteworthy  that the scaling known  for the case of s-wave scattering only,  i.e. $R_{x},R_{y},R_{z}\propto (N\,a)^{1/5}$ is also valid in presence of dipole-dipole interaction. The right hand sides of equations (\ref{equi1}), (\ref{equi2}) and (\ref{equi3}) are written as the product of two different terms: the first on the left is proportional to $N$ and is a function of the condensate radii; the second is a function  only on ratios of radii. 
Thus the first terms fix the scaling of the radii in the same way as in the usual Thomas-Fermi solution with contact interaction. Therefore all of the condensate radii $R_{i}$  rescaled by $N^{1/5}$ are independent from the number of atoms.

\subsection{Expansion}

The expansion can be studied numerically integrating Newton's equations (\ref{ne}) for the particular case 
\begin{eqnarray} \omega_{x}(t) =0, \;\;\;\;\; \omega_{y}(t) =0, \;\;\;\;\;  \omega_{z}(t) =0\quad \text{for} \quad t \ge 0 \end{eqnarray}
with initial condition $R_{i}(0)$ at $t=0$ the initial equilibrium condensate radii.
We write below a general set of equations of motion for the rescaled variables \begin{equation} b_{i}(t) = R_{i}(t)/R_{i}(0) \end{equation}
These are given by
\begin{widetext}
\begin{eqnarray} \label{scaleddynamics} 
{\mathrm{d}^{2}\,b_{x} \over \mathrm{d} t\,^{2}} &=& -   \omega_x^{2}(t) \, b_x +{\bar{\omega}_x^{2}\over b_x^{2} b_{y} b_{z}}
\left[ 1-  \varepsilon_{dd}\, f \left({b_{y} \over b_{x}}\,\kappa_{y}^{0}, {b_{z}  \over b_{x}}\,\kappa_{z}^{0}\right)
-  \varepsilon_{dd}\, b_{x} {\partial f \over \partial b_{x}} 
\left({b_{y} \over b_{x}}\,\kappa_{y}^{0}, {b_{z}  \over b_{x}}\,\kappa_{z}^{0}\right) \right]
\nonumber\\
{\mathrm{d}^{2}\,b_{y} \over \mathrm{d} t\,^{2}} &=& -   \omega_y^{2}(t) \, b_y +{\bar{\omega}_y^{2}\over b_x b_{y}^{2} b_{z}}
\left[ 1-  \varepsilon_{dd}\, f \left({b_{y} \over b_{x}}\,\kappa_{y}^{0}, {b_{z}  \over b_{x}}\,\kappa_{z}^{0}\right)
-  \varepsilon_{dd}\, b_{y} {\partial f \over \partial b_{y}} 
\left({b_{y} \over b_{x}}\,\kappa_{y}^{0}, {b_{z}  \over b_{x}}\,\kappa_{z}^{0}\right) \right] \nonumber\\
{\mathrm{d}^{2}\,b_{z} \over \mathrm{d} t\,^{2}} &=& -   \omega_z^{2}(t) \, b_z +{\bar{\omega}_z^{2}\over b_x b_{y} b_{z}^{2}}
\left[ 1-  \varepsilon_{dd}\, f \left({b_{y} \over b_{x}}\,\kappa_{y}^{0}, {b_{z}  \over b_{x}}\,\kappa_{z}^{0}\right)
-  \varepsilon_{dd}\, b_{z} {\partial f \over \partial b_{z}} 
\left({b_{y} \over b_{x}}\,\kappa_{y}^{0}, {b_{z}  \over b_{x}}\,\kappa_{z}^{0}\right)
\right] \nonumber\\
\end{eqnarray}
\end{widetext}
where $\kappa_{y}^{0}=R_{y}(0) / R_x(0)$ and $\kappa_{z}^{0}=R_{z}(0) / R_x(0)$ are condensate aspect ratios at equilibrium at $t=0$.
Here 
\begin{eqnarray} 
          \bar{\omega}_{i}^{2} = { 15 \, N \hbar^2 a \over  m^{2} \, (R_i)^{2} R_x R_{y} R_z }\,,
\end{eqnarray}
with $i=x,y,z$.
From these equations, we see that, as for the case of contact interaction, the condensate radii once rescaled obey equations that are independent from the number of atoms for a general time-dependent harmonic confinement. The scaling is valid during the dynamics as the time scale of any evolution is uniquely fixed by the trap frequencies \cite{Giovanazzi:2003a}. The scaling properties have important practical consequences as the number of condensate atoms is a difficult parameter to control
experimentally.

\begin{figure*} \centering
\begin{minipage}[t]{0.475\linewidth} \centering \includegraphics[width=70mm]{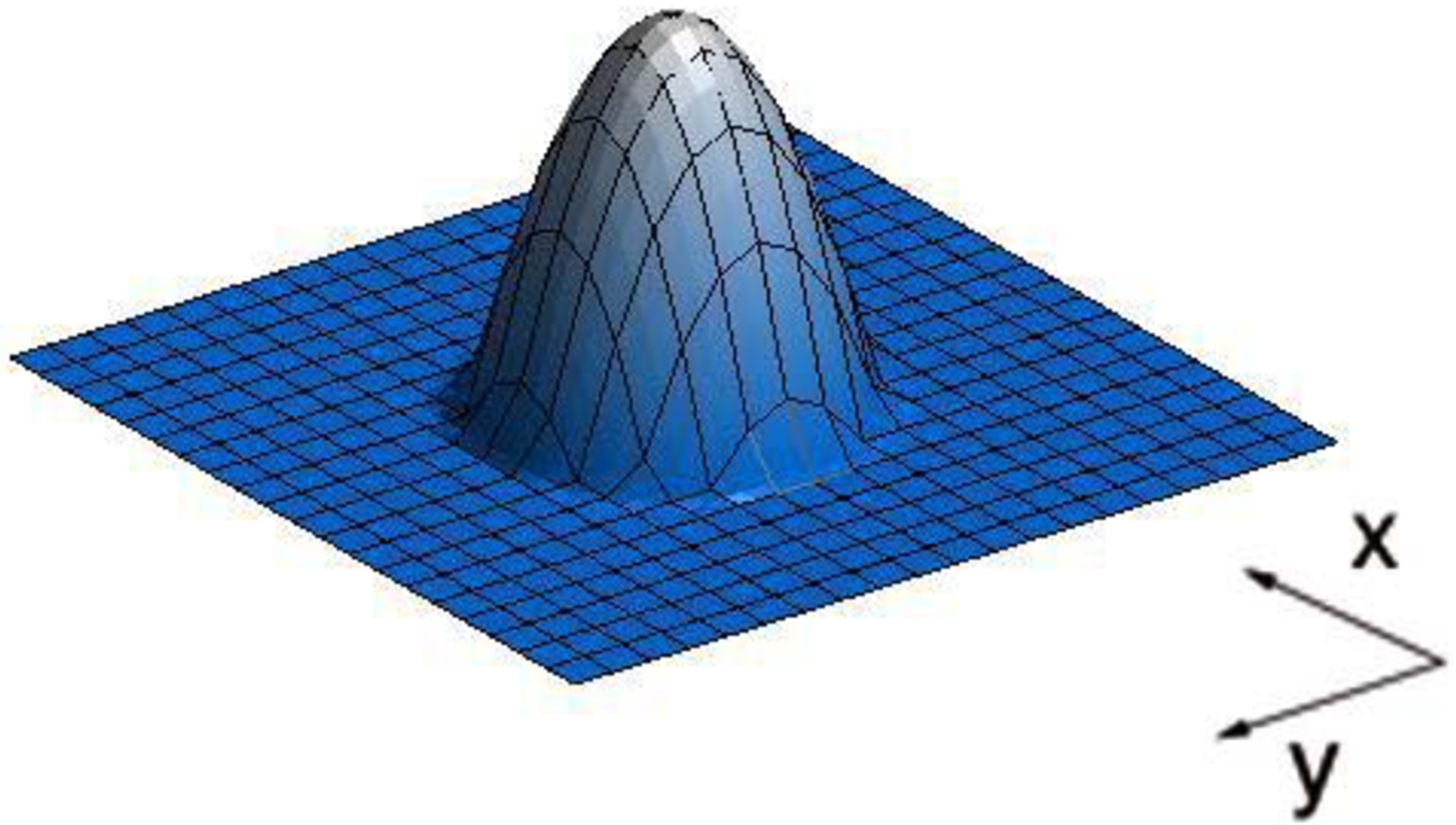} \caption{\label{fig:diptheo:tfprofile}
(Color) Inverted parabolic profile of a BEC in the Thomas-Fermi limit without dipole-dipole interaction. When the trapping potential is symmetric the distribution is also spherically symmetric.} \end{minipage}
\hfill
\begin{minipage}[t]{0.475\linewidth} \centering \includegraphics[width=70mm]{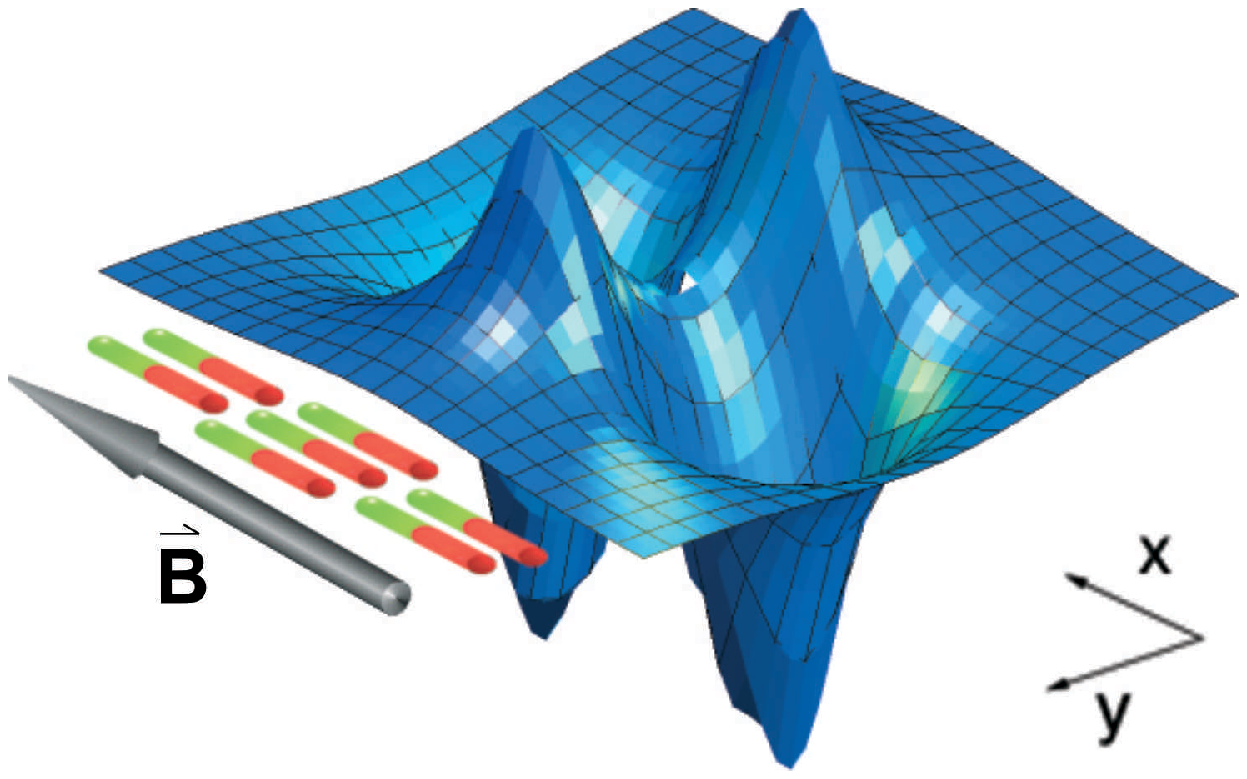} \caption{\label{fig:diptheo:dipolpot}
(Color) Saddle shaped dipole potential generated by dipolar atoms of a BEC in a spherical trap. The atomic dipoles which are illustrated as small magnets in the Figure are aligned by an external magnetic field B. Note the orientation of the
saddle potential relative to the magnetic field direction.}
\end{minipage} \end{figure*}

Before discussing the physics of an expanding dipolar condensate, we shall briefly discuss the expansion dynamics in the case of only contact interaction present (see, e.g. \cite{Castin1996a,BEC_review}). The density distribution of the trapped condensate has the shape of an inverted
paraboloid reflecting the trap anisotropy, e.g. for an isotropic trap the density distribution is also symmetric as depicted in Fig.~\ref{fig:diptheo:tfprofile}. When the condensate is released from the trap the only force acting on the condensate atoms in the Thomas-Fermi limit is (apart from the homogeneous gravity)  equal to minus the gradient of the s-wave scattering mean-field potential, which is proportional to the gradient of the condensate density. Thus the acceleration is stronger in the directions of stronger confinement of the condensate. Since the aspect ratio of the expanded condensate is asymptotically equal to the ratio of the rate of which the condensate radii expand, an inversion of the aspect ratio is generally expected. Therefore, a cigar-shaped condensate results in a pancake-shaped condensate after long times of expansion and {\it viceversa} (see also Figure 3).

\begin{figure*} \centering \includegraphics[width=100mm]{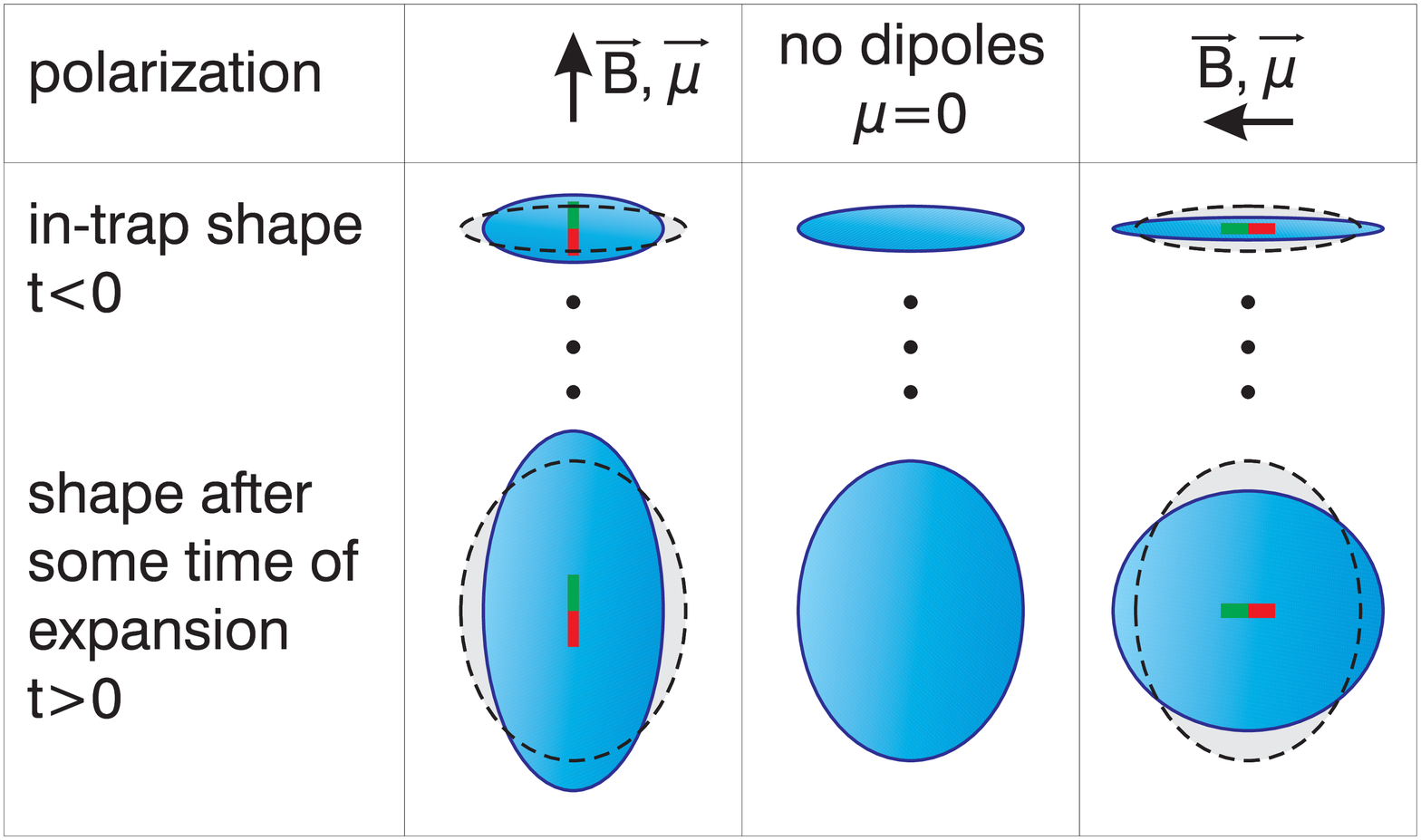} \caption{\label{fig:diptheo:schemexp} (Color) Figure to illustrate the
change of the condensate shape under the influence of magnetization in-trap (top row) and during time of flight (bottom row). Left
column: magnetization in transversal direction; center column: non-dipolar atoms; right column: longitudinal polarization. Dashed ellipses represent the non-dipolar condensate.} \end{figure*}

The effect of dipolar interaction on the condensate expansion can be a little counterintuitive. As already discussed, a trapped cigar-shaped condensate oriented in the direction of the magnetic field has a more pronounced cigar shape before expansion. On the other hand, the more pronounced cigar shape does not manifest in a more pancake shape after the condensate expansion (see Figure 5) as one would expect from the experience of the expansion without dipolar forces. On the contrary, the expanded condensate has a less pancake-like shape.  The general trend of deforming the condensate with an elongation along the magnetization and a contraction in the transversal directions is kept also during the expansion of the condensate as can be seen from the schematic drawing of Figure~\ref{fig:diptheo:schemexp}. Let us discuss that in more details.

To get insight into this behaviour, we should look at the dipole-dipole mean-field potential $\Phi_{dd}$. The general expression of $\Phi_{dd}$ is given in appendix B. Its main characteristics relevant for the present discussion are contained in the special case of a spherical symmetric condensate with radius $R_{TF}$ (see Figure 2).  For $r<R_{TF}$, the dipole-dipole mean-field potential at a position $\vec{r}$ with distance $r$ from the center of mass is given by~\cite{Giovanazzi:2002a} 
\begin{equation}
\Phi_{dd}(\vec{r})=\frac{\varepsilon_{dd} m
\omega_0^2}{5}\left[1-3\left(\frac{\vec{e}_{\mu}\cdot
\vec{r}}{|r|}\right)^2\right]r^2\quad \text{for} \quad r\leq R_{TF}.
\end{equation}
From this equation, it becomes clear that the potential is harmonic in $r$ but has an angular dependence. Note that the term in square brackets varies depending on the angle between the orientation of the dipoles $\vec{e}_{\mu}$ and the position vector $\vec{r}$ between $-2$ if the position and the polarization are parallel and $+1$ if they are orthogonal. The potential therefore has the form of a saddle with a negative curvature along the direction of magnetization and a positive curvature in transverse direction.

After a sudden switch-off of the external potential, the only forces stem from the gradient of the contact mean-field potential (repulsive in all directions) and the gradient of the dipole-dipole mean-field interaction (repulsive and attractive). Due to its direct proportionality to the local density, the contact part of the mean-field potential reveals the same parabola shape as depicted in Fig.~\ref{fig:diptheo:tfprofile}. 
The dipole-dipole potential $\Phi_{dd}(\vec{r})$ still has its (harmonic) saddle shape (see Figure~\ref{fig:diptheo:dipolpot}). 
Note that in the direction of magnetization, the gradient of the potential energy of the total mean field ($U_{mf}=g|\phi(\vec{r})|^2+\Phi_{dd}(\vec{r})$) will be larger than without dipole-dipole interaction. Therefore the atoms will obey a larger acceleration along the direction of magnetization than without dipole-dipole interaction. In the directions perpendicular to the magnetization, the condensate atoms attract each other.  Thus the repulsive contact interaction is weakened by the dipole-dipole interaction in transversal direction and the acceleration that atoms feel perpendicular to the magnetization will be smaller. This explains why the general trend of deforming the condensate with an elongation along the magnetization and a contraction in the transversal directions is kept also during the expansion of the condensate.
\begin{figure*}[htb] \centering \includegraphics[width=130mm]{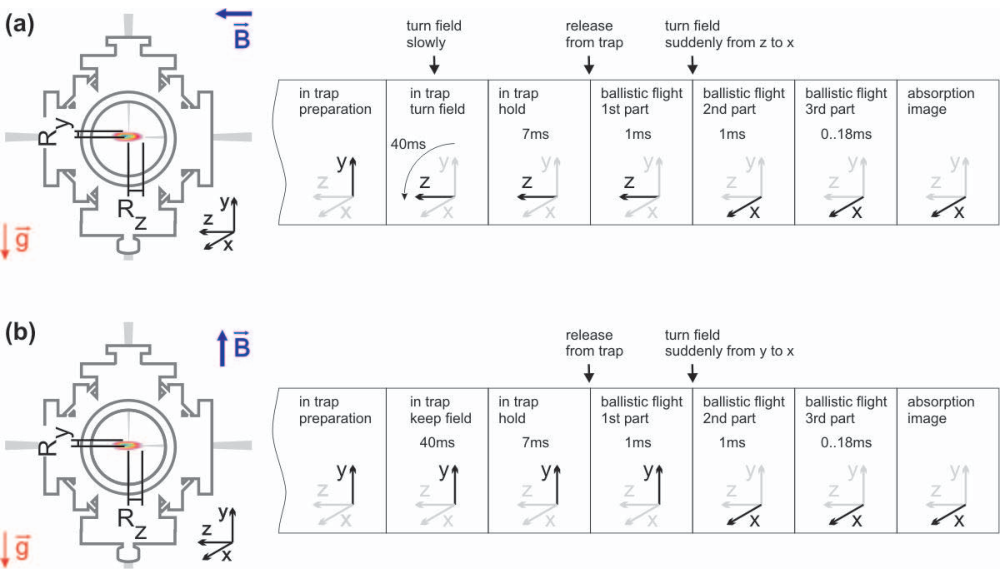} \caption{\label{fig:dipexp:experiment} (Color) Experimental cycle for
measuring the dipole-dipole interaction. Left Figures: alignment of the field relative to the chamber just before releasing them. Gravity $\vec{g}$ marks y as the vertical axis. Right Figures: schematic cycle. The magnetic field during preparation is in both cases along the $y$-axis. To measure with $z-$polarization (a), the field has to be turned slowly (within \unit[40]{ms}) within the trap before releasing the atoms. After \unit[1]{ms} of expansion, the field is switched to $x$-direction. For $y-$polarization (b), the field is kept in $y-$direction until \unit[1]{ms} after release.} \end{figure*}

\section{Experiment}

To measure the effect of the magnetic dipole-dipole interaction on the expansion dynamics of the condensate, we prepare a BEC in a crossed optical dipole trap. The details of the preparation of the chromium BEC are discussed in \cite{Griesmaier:2006a}. 
By decreasing the intensity of the horizontal trapping beam from initially \unit[9.3]{W} to \unit[280]{mW} and keeping the vertical intensity constant at \unit[4.5]{W}, we produce almost pure condensates containing on average a number of $40000$ atoms. A schematic illustration of the subsequent experimental cycle is depicted in Figure~\ref{fig:dipexp:experiment}. After \unit[250]{ms} equilibration, the intensity of the horizontal beam is increased adiabatically to \unit[2.3]{W} to form an anisotropic trap (trap parameters $f_x =$\unit[942]{Hz}, $f_y=$\unit[712]{Hz}, and $f_z=$\unit[128]{Hz}). Throughout the preparation procedure, we keep a homogeneous offset field of $\sim$\unit[11.5]{G} along the $y-$axis until the trap has been ramped up \cite{Griesmaier:2006a}. 
After this change of the trap parameters, we either keep the field aligned along $y$ (situation b) in Fig.~\ref{fig:dipexp:experiment}) or we rotate the field adiabatically from the $y-$ to the $z-$direction (situation a) in Fig.~\ref{fig:dipexp:experiment}). This is done by increasing the field in $z-$direction linearly within \unit[40]{ms} to $\sim$\unit[11.5]{G} while reducing the field in $y-$direction during the same time to \unit[0]{G}. After the field has reached the steady state, we keep the atoms for another \unit[7]{ms} in the trap to give them enough time to redistribute. The total storage time in both cases of longitudinal ($z$) or transversal ($y$) magnetization is equal. Subsequently, the atoms are released by a sudden switch-off of the trapping beams. The polarization field is kept constant for \unit[1]{ms} after release from the trap and then rotated quickly to the transversal $x-$axis in either case by switching on the field in the $x-$direction and switching off the $z-$ or $y-$ fields. The field along  $x$ is needed to align the atomic magnetic moments for absorption imaging in this direction. The \unit[1]{ms} of free expansion before switching the field is long enough for the mean-field energy to drop to such a small part of its initial value that changing the alignment of the dipoles after this time does not influence the expansion anymore. In other words, after this time, the gas is already so dilute that any kind of interaction among the atoms can be neglected compared to the kinetic energy. After an additional time of flight of up to \unit[18]{ms} plus \unit[1]{ms} for the detection field to settle (total time of flight \unit[2]{ms} to \unit[20]{ms}), an absorption image of the cloud is taken.\\
\begin{figure*} \begin{minipage}[t]{0.475\linewidth} \includegraphics[width=80mm]{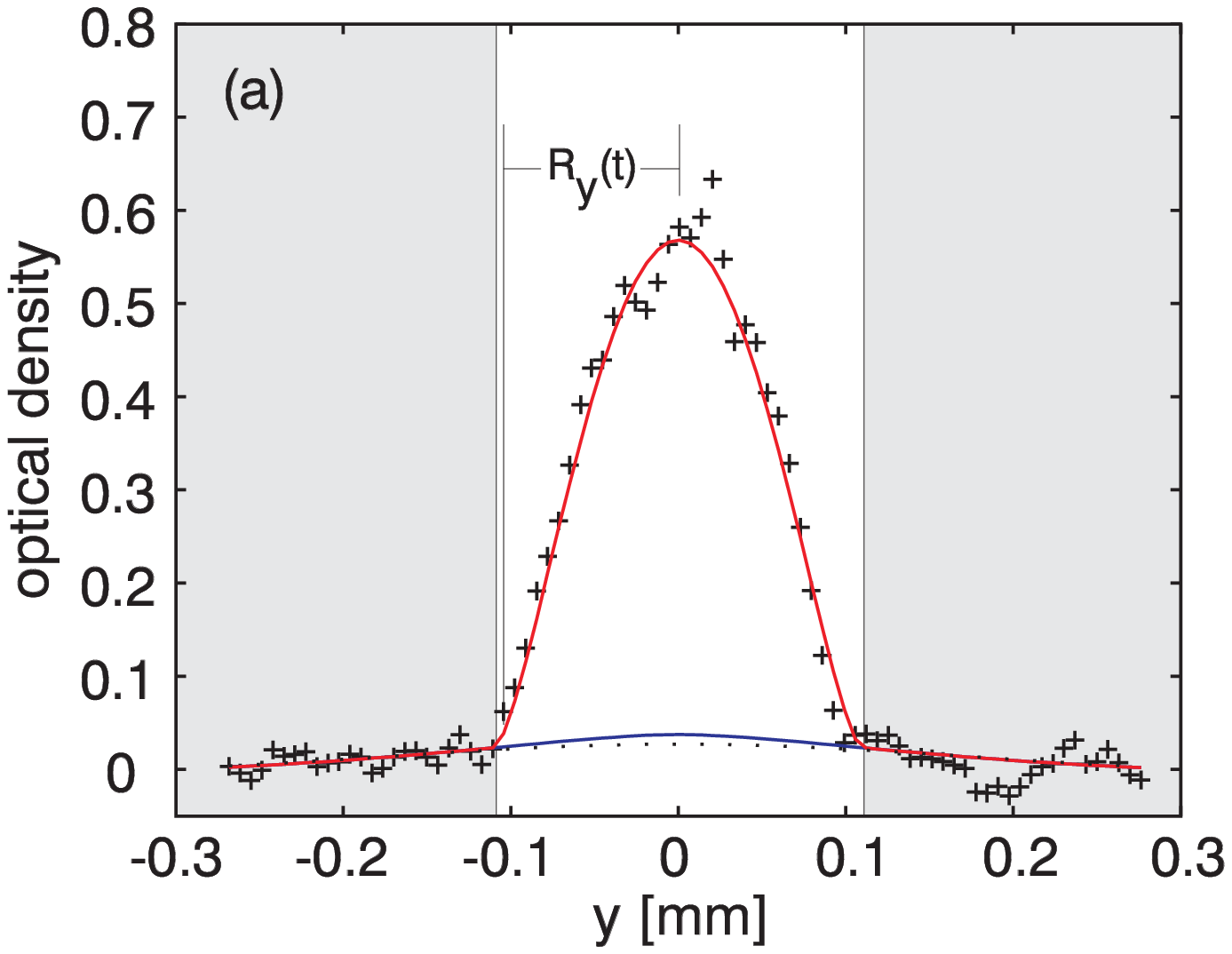} \end{minipage}
\hfill \begin{minipage}[t]{0.475\linewidth} \includegraphics[width=80mm]{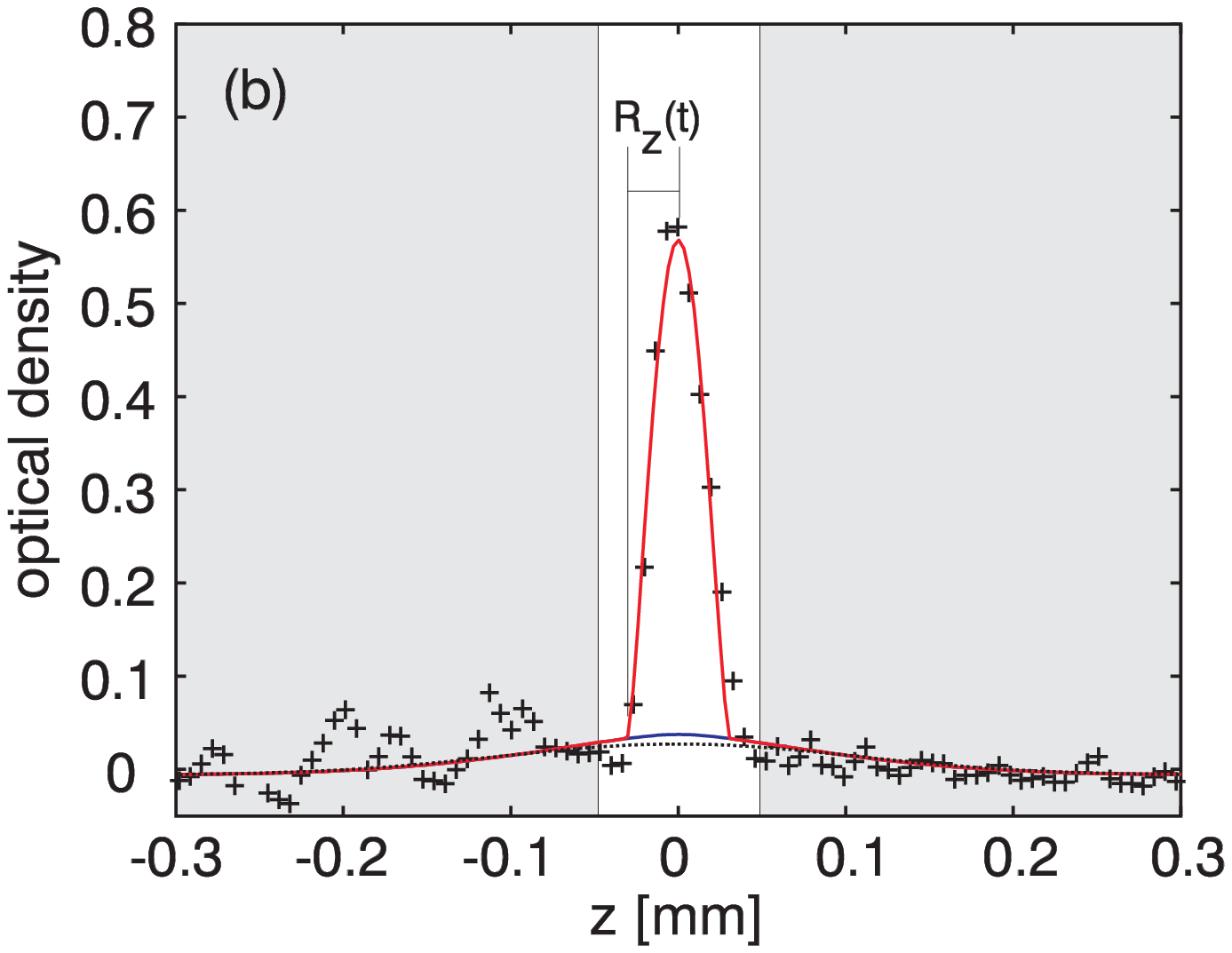} \end{minipage} \caption{\label{fig:dipexp:profiles} (Color) Density profiles in y- and z-direction of an almost pure condensate  after \unit[90]{ms} of expansion. Grey shaded areas have been excluded from the fit of the condensate cloud.} \end{figure*}

The images were evaluated by two dimensional fits to the density profiles. Figure~\ref{fig:dipexp:profiles} shows 1D cuts through the density profile of an expanded, almost pure condensate in the $y$- and $z$- direction, respectively. The most convenient quantity to analyse the expansion is the aspect ratio $\kappa=R_y/R_z$ since it is insensitive to fluctuations of the number of atoms. The only quantities that have to be known exactly are the trap parameters and the ratio $\varepsilon_{dd}$ between magnetic dipole-dipole interaction and contact interaction (\ref{epsilon}). The trap frequencies have been determined using a parametric heating technique \cite{Griesmaier:2006a}.

\begin{figure*} \centering \begin{minipage}[t]{0.600\linewidth} \includegraphics[width=1\columnwidth]{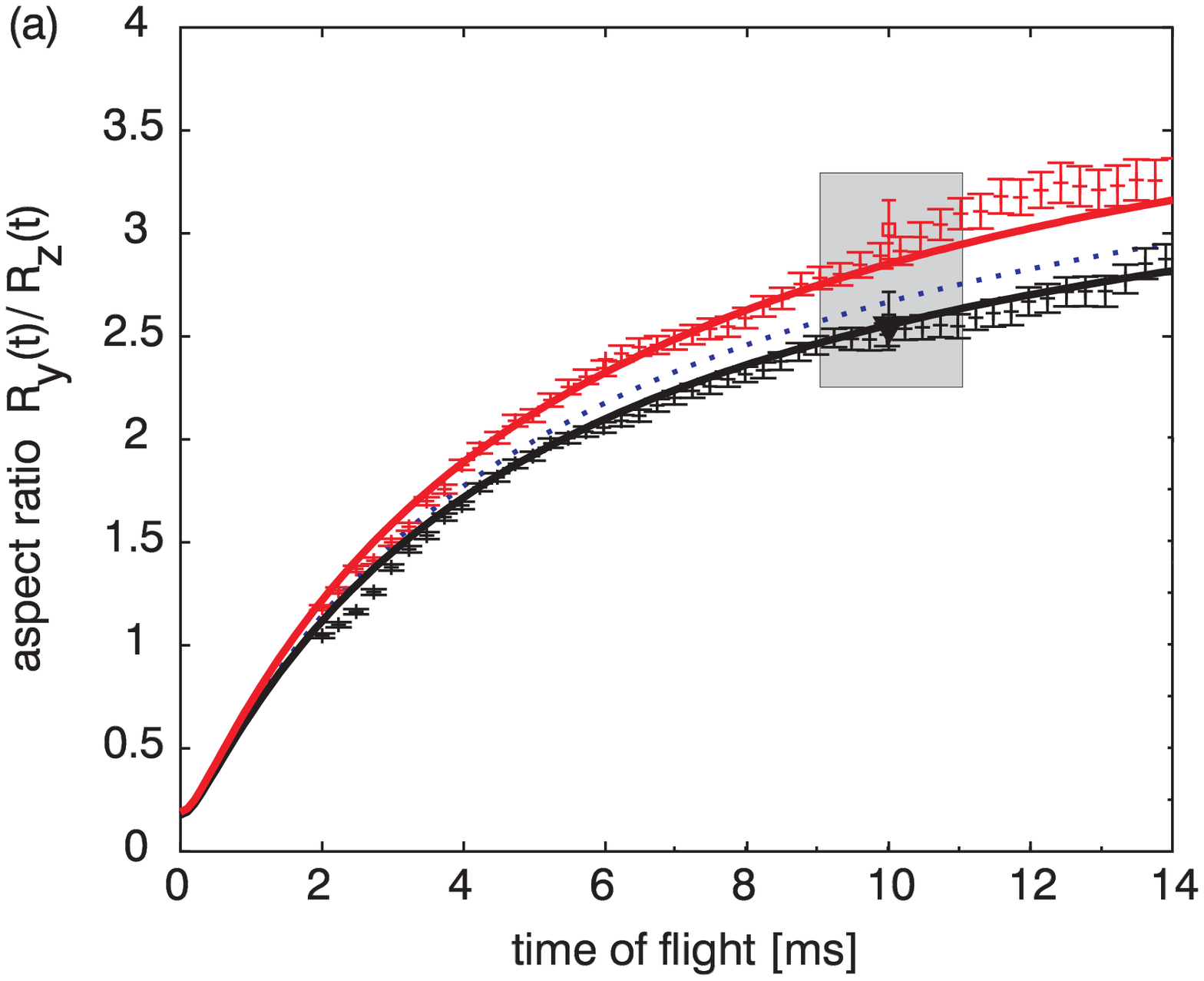}
\end{minipage} \begin{minipage}[t]{0.091\linewidth} \includegraphics[width=1\columnwidth]{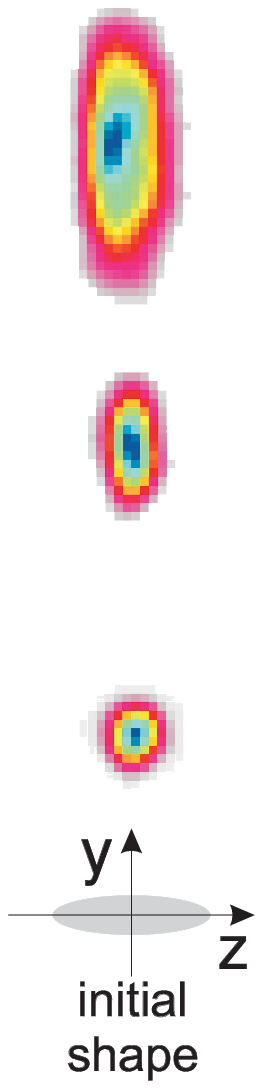} \end{minipage}
\hfill \begin{minipage}[t]{0.600\linewidth} \includegraphics[width=1\columnwidth]{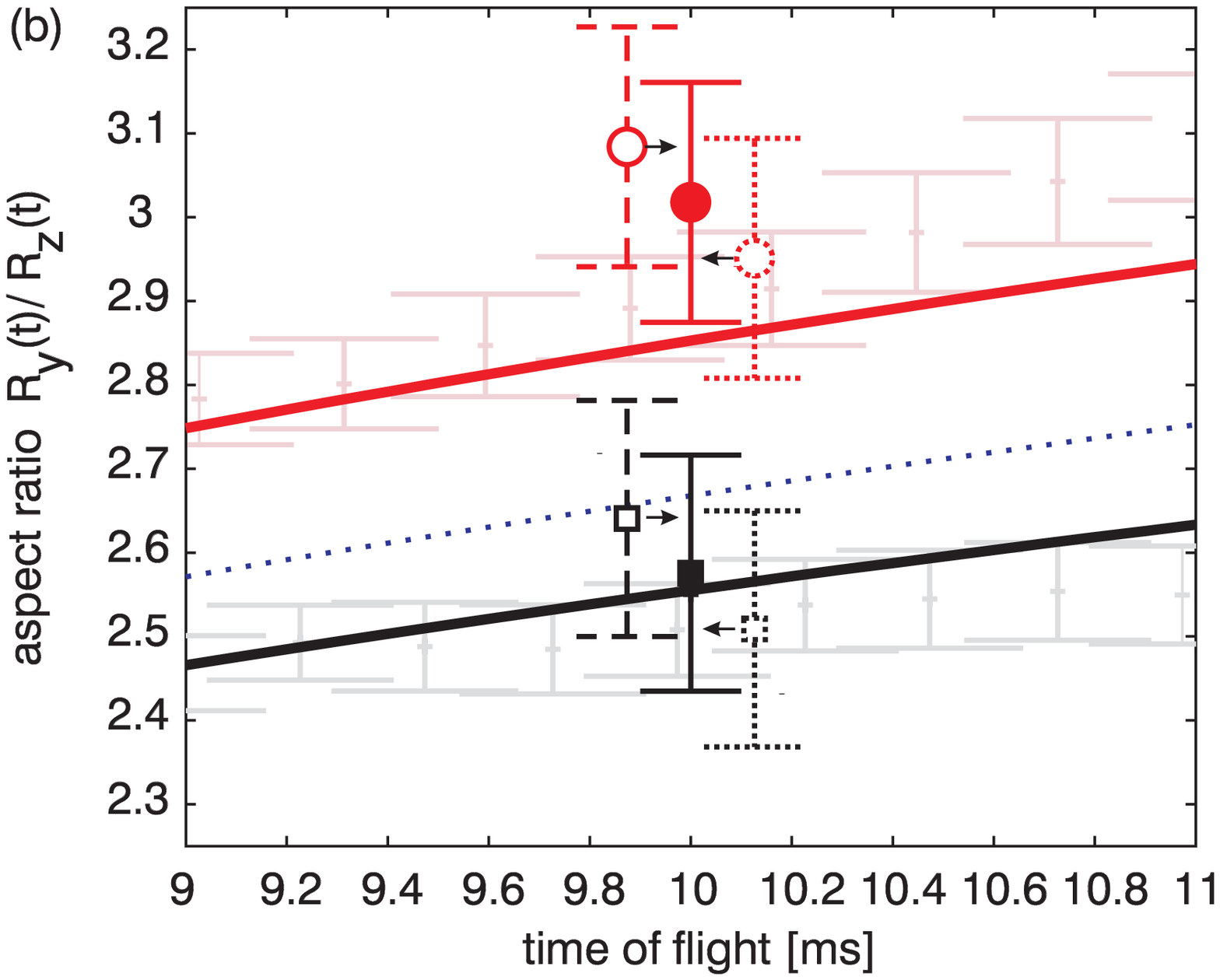}
\end{minipage} \begin{minipage}[t]{0.091\linewidth} \includegraphics[width=1\columnwidth]{empty} \end{minipage} 
\caption{\label{fig:dipexp:data} \label{fig:evap:dipolarexp}
(Color) Aspect ratio of the expanding dipolar condensate. Data points in Figure (a) are averaged with an 11-point moving average. Error bars in the upper graph represent errors from the fits to the density distribution. Upper, red data: Field aligned in vertical y-direction. Lower, black data: Field in horizontal z-direction. The solid lines represent the corresponding theoretical predictions. The blue dotted line is the behavior that one would expect for pure contact interaction without the presence of dipole-dipole interaction among the atoms. Figure (c) illustrates the evolution of the condensate shape as seen by the camera. Figure (b) shows in details the grey shaded area in Figure (a). At ten milliseconds time of flight, we have performed a series of measurements under the same conditions. The two data points
represent the mean values of 42 measurements with y-polarisation (red circle) and 32 with z-polarisation (black square). The solid error bars are derived from the statistics of the measured value and represent one standard deviation in both directions. They do not include the systematic error on the single measurements. These errors affect both the measurements with z- and y-polarisation in the same way and do not change the significance of the measured difference in the expansion dynamics with different polarisation. The shift of the measured aspect ratios due to such systematic errors is indicated by the dashed/dotted representation of the error-bars (displaced laterally for clarity) in Figure (b).} \end{figure*} 
Figure~\ref{fig:dipexp:data} shows the aspect ratio of the BEC for different times of ballistic expansion. The set of data marked with red squares represent the experiments performed with polarization in vertical ($y-$) direction and black circles represent the results
with horizontal ($z-$) polarization. The upper graph shows the result of sequential experiments were the total time of flight was
varied between \unit[2]{ms} and \unit[14]{ms}. Since one run (i.e. catching atoms in the CLIP trap, Doppler cooling, compressing, rf-cooling, transfer to the ODT, pumping, plain evaporation in the ODT, forced evaporation, modification of the trap and the fields,
taking an image, resetting the system), takes about \unit[1]{minute} and \unit[20]{seconds}, the data of the time of flight series presented in Figure~\ref{fig:dipexp:data} corresponds to a total measuring time of more than \unit[4]{hours}. To reduce the influence of systematic drifts during that time, the time of flight of subsequent pictures was chosen randomly. For the same reason, we also changed between $y-$ and $z-$ polarization every $10$ runs. An $11$-point linear moving average (corresponding to averaging over \unit[2.2]{ms} in the Figure) has been applied to both sets of data in the left graph to average out fluctuations in the determined condensate widths. A moving average of that length is reasonable since the expected behavior does not show features on shorter time scales that could be concealed by the averaging. This has been
proven by applying the same moving average to the theoretical values. To be able to display also all the measured data for short times of flight, the range of the moving average was increased from $1$ to $11$ within the first six data points. The data point corresponding to \unit[2]{ms} time of flight is thus not averaged, the one at \unit[2.2]{ms} is averaged over $3$ points, the one at \unit[2.4]{ms} over $5$, and so on. Thus, only the data corresponding to \unit[2]{ms} to \unit[2.8]{ms} TOF are avereaged over less than $11$ points.\\

\subsection{Comparison of experiment and theory}

The measured data for the condensate aspect ratio are compared to the numerical results obtained integrating Newton's equations (\ref{ne}) in Figure 6. The theory contains no adjustable parameters. It only relies on known or measured quantities, namely the trap frequencies, the magnetic moment, and the s-wave scattering length that characterises the contact interaction~\cite{Werner:2005}. The  dotted line represents the expectation for a gas interacting solely via s-wave scattering. Compared to this non-dipolar behavior, the expansion of the condensate shows a dependence on the polarization of the atoms that is in agreement with the theory: With transversal polarization (field along the $y-$axis), the condensate is elongated in the transversal direction and the aspect ratio is increased; if the polarization is in longitudinal direction (field along the $z-$axis), the condensate is contracted in vertical direction and the aspect ratio is decreased. Also the quantitative agreement is remarkable.\\

The error bars in the first graph of Figure~\ref{fig:dipexp:data} include only errors that stem from the fit of the condensate size. Systematic errors, e.g. uncertainty of the magnification are not contained. These systematic errors can be found in the lower graph of Figure~\ref{fig:dipexp:data} where the mean value of the results of $42$ and $32$ measurements with $y-$ and $z-$ polarization, respectively, are presented. All these  measurements were performed after the same time of flight of \unit[10]{ms}. In all, $50$ measurements have been performed at \unit[10]{ms} with both polarizations. However, some of the measurements had to be withdrawn due to an obvious instability of the system which on the one hand lead to a number of shots where the number of atoms was substantially smaller than the average $(4.0\pm0.6)\;\cdot10^4$ of the remaining measurements. On the other hand, some of the condensates did not fall down vertically but moved significantly to one or the other side during their flight, which we considered as a signature that the condensate was kicked (probably by mechanical noise on the optical table) prior to or when switching of the trap and also these images were withdrawn. The error bars directly connected to the two data points are the statistical errors of all measurements and represent $\pm 1$ standard deviation from the mean value. The  error bars on the left represent the systematic errors, that are additionally contained in the data, assuming a systematic \unit[2]{\%} uncertainty in the size of the cloud. Note that such an error is contained in all data points in the same way and would not change the relative difference between the expansion data for the two polarizations. Taking also this systematic error into account, also the upper data point for $y-$polarization, which deviates a little from the theoretical expectation, is also within error bars with respect to the theoretical prediction obtained with  $ \epsilon_{dd}^{Cr}\approx 0.15 $. A finer investigation of the agreement between the theory and the experiment will be the subject of a further work.

\section{Conclusion}

The observed mechanical manifestation of dipole-dipole interaction in a Bose-condesate gas of chromium is in very good quantitative agreement with the theory of dipolar gases in the Thomas-Fermi limit. The trapped condensate atoms redistribute depending on the direction of the applied magnetic field. Similar to what occurs with magnetic solid particles or liquids (ferrofluids), the strongly magnetic chromium atoms align preferably along the direction of magnetization. This induces a change in the shape of the condensate. The micron-sized chromium condensate needs to be released from the trap in order to be imaged after expansion. The induced change in shape remains visible also after release of the condensate from the trap and are quantitatively well described within the framework of a generalized Gross-Pitaevskii equation, which proves to be an appropriate description.

The chromium Bose-Einstein condensate opens fascinating perspectives for the experimental study of dipole-dipole interaction induced magnetism in gaseous systems. Since one can exploit Feshbach resonances~\cite{Werner:2005} to adjust contact-like (isotropic and short-range) atom-atom interactions and use rotating magnetic fields to tune the dipole-dipole interaction~\cite{Giovanazzi:2002a}, interaction regimes ranging from only contact to purely dipolar can be realized. Depending on the relative strengths of these two interactions and on the absolute strength of the dipole-dipole interaction, many exciting phenomena are expected.

\begin{acknowledgments}
We acknowledge support by the Landesstiftung Baden-W{\"u}rttemberg, the Alexander von Humboldt Foundation, the German Science Foundation (DFG) (SPP1116 and SFB/TR 21), the 
Minist{\`e}re da la Recherche (grant ACI Nanoscience 201), the ANR (grants NT05-2-42103 and 05-Nano-008-02), and the IFRAF Institute.
\end{acknowledgments}

\appendix

\section{Dipole-dipole mean-field energy}

The mean-field magnetic dipole-dipole energy is given by
\begin{eqnarray} H_{dd}^{x}&=& - {15\over 7} \left({ N^2 \hbar^2 a \over  m }\right)  {\varepsilon_{dd}\,f(\kappa_{yx},\kappa_{zx})\over R_x R_{y} R_z} \label{ahddx} \end{eqnarray} where the magnetic field is assumed parallel to the $\hat{x}$ axes and $\kappa_{yx}=R_y/R_x$ and $\kappa_{zx}=R_z/R_x$  are condensate aspect ratios.
The function $f$ is given by 
\begin{eqnarray}\label{affunction} f(\kappa_{yx},\kappa_{zx})&=&1+3\kappa_{yx}\kappa_{zx} 
{\mathrm{E}(\varphi\setminus\alpha) -\mathrm{F}(\varphi\setminus\alpha) \over (1-\kappa_{zx}^{2})\sqrt{1-\kappa_{yx}^{2}}}
\end{eqnarray} with 
\begin{eqnarray} \sin\varphi&=&\sqrt{1-\kappa_{yx}^{2}}\\ \sin^{2}\alpha&=&{1-\kappa_{zx}^{2}\over1-\kappa_{yx}^{2}}
\end{eqnarray}
Here $\mathrm{F}(\varphi\setminus\alpha)$  and  $\mathrm{E}(\varphi\setminus\alpha)$ are the incomplete  elliptic integrals of the first and second kinds \cite{Abramowitz:}. 
\begin{figure}[hhh!]  \includegraphics[width=80mm]{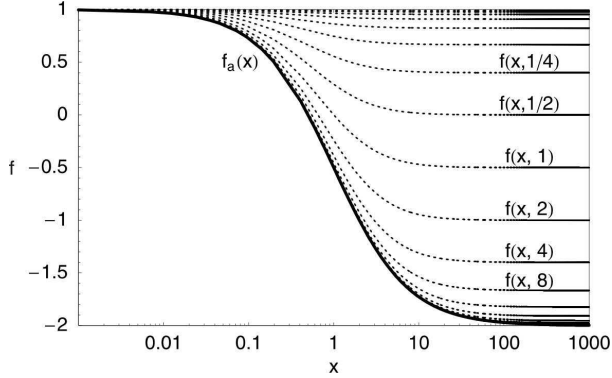} \caption{\label{figure1}
Log-linear plot of the function $f(x,y)$ defined in equation (\ref{affunction})  versus $x$ and for different values of  $y$ (dashed lines). The chosen values of $y$ correspond to different powers of 2. 
Due to symmetry property $f(x,y)=f(y,x)$ of the function $f$ the asymptotic values of $f(x,y)$ for large $x$ correspond to the function $f_{a}(y)$ defined in (\ref{fafunction})  which is also shown as function of $x$ (solid line).  }
\end{figure} 
Figure 7 and 8 show the typical behavior of the function $f$, which is symmetric 
\begin{eqnarray}  f(x,y) =f(y,x) \end{eqnarray} and is a smooth and limited function with the property that 
\begin{eqnarray} 1 \ge f(x,y) \ge -2 \end{eqnarray}
For small values of one of its arguments $f$ is equal to 1
\begin{eqnarray} f(x,0) = 1 \end{eqnarray}
When one of the argument is very large its asymptotical values are given  by the following function
\begin{eqnarray}\label{fafunction} f_{a}(x)&=&f(\infty,x)\nonumber\\&=&  1-3{(1-x)x \over 1-x^{2}} \end{eqnarray}
with the property \begin{eqnarray} f_{a}(x)+f_{a}(1/x)+1=0 \end{eqnarray}
When both arguments are very large
\begin{eqnarray} f(\infty,\infty) = -2 \end{eqnarray}
In the special case of equal arguments, $f$ becomes a function of only one variable and is given by
\begin{eqnarray}
f_{s}(\kappa)&=& f(\kappa,\kappa)\nonumber\\&=&
\frac{1+2\kappa^2}{1-\kappa^2}-\frac{3\kappa^2
{\tanh}^{-1}\sqrt{1-\kappa^2}}{\left(1-\kappa^2\right)^{3/2}}
\end{eqnarray}
which was already introduced considering the special case of cylindrical symmetric condensate \cite{Giovanazzi:2003a, ODell:2004a, Eberlein:2005}. Moreover, $f$ obeys the sum rule
\begin{eqnarray} f(x,y)+f\left( { y \over x}, {1\over x}\right) +f\left( { 1 \over y}, {x\over y}\right) = 0 \end{eqnarray}
This equation has the physical meaning that the average over all  directions of the polarization gives zero contribution to the dipolar energy. Using the above equation, it is possible to show for instance that 
\begin{eqnarray} f(x,1)=-{1\over 2} f_{s} \left( { 1 \over x}\right) \end{eqnarray}
\begin{figure}[htbp] \includegraphics[width=0.49\textwidth]{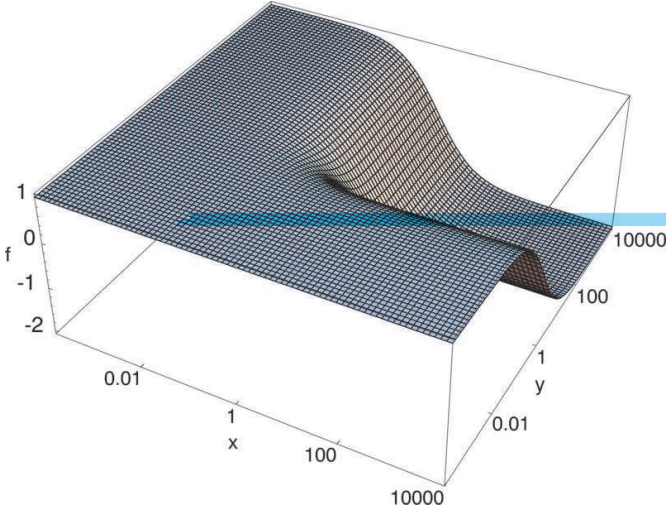}  \caption{\label{figure2}
(Color) Log-linear 3-dimensional plot of the function $f(x,y)$ defined in equation (\ref{affunction}) versus $x,y$. } \end{figure}
With the above relationship, it is easy to calculate a polynomial expansion of $f$ around the point $(x,y)=(1,1)$ which represents a spherical condensate. We give below the expression of a cubic approximation that can be useful for aspect ratios of order of unity
\begin{eqnarray} &\; & f(1+x,1+y) \approx \frac{-2\,(x+y)}{5} 
 +\frac{9\,(x^2+y^2)-8\,x\,y}{35}  \;\;\; \nonumber\\ &\; &
 +\frac{12\,\left( x^2\,y + x\,y^2 \right) -16\,\left( x^3 + y^3 \right)}{105} 
\end{eqnarray}
 For aspect ratios in the range of $(0.5,1.6)$ the absolute error of the above polynomial approximation $f_{approx}$ is given by $0<f - f_{approx}<0.02$.

\section{The mean-field dipole-dipole potential}

The magnetic dipole-dipole contribution to the mean-field potential
for dipoles polarized in the $\hat{x}$ direction is given by 
\begin{eqnarray}
&\,&
V_{\mathrm{mf}}^{x} =
{ 45 \varepsilon_{dd}  N \hbar^2 a \over  2 m R_x R_y R_z}
\left(-{1\over3}f
      +f_{x}{x^{2}\over R_x^{2}}
      +f_{y}{y^{2}\over R_y^{2}}
      +f_{z}{z^{2}\over R_z^{2}}
      \right)\;\;\;\;\;\;\;
\label{vmfx}
\end{eqnarray}
where $\{f,f_{x},f_{y},f_{z}\}$ are function of the aspect ratios $\{\kappa_{yx},\kappa_{zx}\}$ and are given below
\begin{widetext}
\begin{eqnarray}
&\,& 
f_{x}(\kappa_{yx},\kappa_{zx})={1\over 3}
-      { \kappa_{yx}^{2} \kappa_{zx}^{2} 
       \over
       (1-\kappa_{yx}^{2})  (1-\kappa_{zx}^{2})   }
+2     {   \kappa_{yx} \kappa_{zx} 
       \over
       \sqrt{1-\kappa_{yx}^{2}}  (1-\kappa_{zx}^{2})^{2}   }
\left( 1+ {1-\kappa_{zx}^{2} \over 1-\kappa_{yx}^{2}} \right)
\mathrm{E}(\varphi\setminus\alpha)
\nonumber\\&\,&
-     {  \kappa_{yx} \kappa_{zx} 
       \over
       \sqrt{1-\kappa_{yx}^{2}}  (1-\kappa_{zx}^{2})^{2}   }
\left( 2+ {1-\kappa_{zx}^{2} \over 1-\kappa_{yx}^{2}} \right) \mathrm{F}(\varphi\setminus\alpha)
\\
&\,& 
f_{y}(\kappa_{yx},\kappa_{zx}) = {1\over 3}
+      { \kappa_{yx}^{2} \kappa_{zx}^{2} 
       \over
       \kappa_{yx}^{2} - \kappa_{zx}^{2} }
+      { \kappa_{yx}^{4} \kappa_{zx}^{2} 
       \over
       (1-\kappa_{yx}^{2})  (\kappa_{yx}^{2} - \kappa_{zx}^{2})   }
+     {   \kappa_{yx}^{3} \kappa_{zx} 
           ( \kappa_{zx}^{2} - \kappa_{yx}^{2} -1  )
       \over
       (1-\kappa_{yx}^{2})^{3/2}  (1-\kappa_{zx}^{2})^{2}   
       (       \kappa_{yx}^{2} - \kappa_{zx}^{2} )}
\mathrm{E}(\varphi\setminus\alpha)
\nonumber\\&\,&
+     {   \kappa_{yx}^{3} \kappa_{zx}  
       \over
       (1-\kappa_{yx}^{2})^{3/2}  (1-\kappa_{zx}^{2})^{2}   }
\mathrm{F}(\varphi\setminus\alpha)
\\ 
&\,& f_{z}(\kappa_{yx},\kappa_{zx}) =f_{y}(\kappa_{zx},\kappa_{yx}) 
\end{eqnarray}
\end{widetext}
Here $\mathrm{F}(\varphi\setminus\alpha)$  and  $\mathrm{E}(\varphi\setminus\alpha)$ are the incomplete  elliptic integrals of the first and second kinds \cite{Abramowitz:}. Their arguments are given by
\begin{eqnarray} \sin\varphi&=&\sqrt{1-\kappa_{yx}^{2}}\\ \sin^{2}\alpha&=&{1-\kappa_{zx}^{2}\over1-\kappa_{yx}^{2}}
\end{eqnarray}
The $\{f,f_{x},f_{y},f_{z}\}$ obeys the identity given by 
\begin{eqnarray} 
f=f_{x}+ f_{y}+f_{z}
\end{eqnarray}

\bibliography{biblio}

\end{document}